\documentclass[11pt]{article}
\usepackage{moriond,epsfig}

\def\be{\begin{equation}}
\def\ee{\end{equation}}
\def\bea{\begin{eqnarray}}
\def\eea{\end{eqnarray}}

\begin{document}
\vspace*{4cm}
\title{PHENOMENOLOGY OF JET PRODUCTION \\
 IN FORWARD DIRECTION AT HIGH ENERGY HADRON COLLIDERS }

\author{ K. Kutak }

\address{Department of Physics, University of Antwerp, Groenenborgerlaan 171,\\
B-2020, Belgium}

\maketitle\abstracts{We calculate observables relevant for forward jets
at LHC. The simulations are performed using Monte Carlo event generators. In particular we compare results
from CASCADE based on high energy factorization and PYTHIA which is based on collinear factorization.}

\section{Introduction}
Experiments at the Large Hadron Collider  (LHC) will allow to test the Standard Model at very high energies. Here we are interested in Quantum 
Chromodynamics (QCD)
processes like forward jet production~\cite{Aslanoglou:2007wv,Jung:2009eq}. 
This process is of particular interest since it will allow for better understanding of partonic structure of the proton at extreme energies. 
The large center of mass energy at the LHC will require application of QCD resummation approaches capable to account for multiple scales 
in the problem~\cite{Stirling:1994zs,DelDuca:1993zw}. Namely, one has to account for logarithms of type $\alpha_s^n\ln^mp_\perp/\Lambda_{QCD}$ where
p$_\perp$ is a transversal momentum of produced jet and 
another type of logarithms: $\alpha_s^n\ln^m1/x$ \cite{Lipatov:1976zz,Kuraev:1977fs} due to the fact that one of the incoming proton will be probed at very small 
longitudinal momentum fraction. The theoretical framework to resumme consistently both kinds of logarithmic corrections  
in pQCD    
is based on  high-energy factorization at fixed transverse momentum~\cite{Catani:1990xk,Catani2,Catani3}. 
This formulation    depends  on  unintegrated parton distributions, obeying appropriate evolution equations, and short-distance, 
process-dependent matrix elements.
The  unintegrated-level evolution is given by  evolution equations in rapidity,   or angle,  parameters.   
Different  forms of the  evolution, valid  in different kinematic regions, are available,
see~\cite{Collins:2008ht,Hautmann:2009zz,Rogers:2008jk,Hautmann:2007uw}, 
and references therein.
In this article we apply recently obtained \cite{Deak:2009xt,Deak:2010gk} results for hard matrix elements relevant for forward jet physics
together with parton shower Monte Carlo generator CASCADE ~\cite{Jung:2001hx,Jung:2010si} for calculating
observables related to forward jet phenomenon. For other approaches to the phenomenon of jets at high energies we refer the reader to the following papers: \cite{Andersen:2009nu,Colferai:2010wu,Andersen:2011hs}. The paper is organized as follows.
In Sec.~2 we recall elements of high energy factorization framework relevant for our study. 
In Sec.~3 we present phenomenological results for jet production focusing on transversal momentum spectra and rapidity spectra. 
\section{Factorization kinematics and matrix elements relevant for forward jets}
Forward jet associated with central jet is a process where after collision of two protons
one collimated group of high p$_\perp$ hadrons continues along the direction of one
of colliding protons - forward detector region, while another group heads toward central region.
The high p$_\perp$ production at  microscopic level can be understood as originating from collision of
two partons where one  of them which is almost on-shell carries large longitudinal momentum fraction $\xi_1 p_1$ of
mother proton ($p_1$) while the other one carries small longitudinal momentum fraction $\xi_2 p_2$
of the other proton ($p_2$) and is off-shell, 
where $k_1$, $k_2$, are the four momenta of initial state partons and $p_3$ and $p_4$ are
four-momenta of final state partons.\\
The framework to describe forward jets is provided by
high-energy factorization which was derived after observation of gluon exchange dominance at high
energies.
Similarly to collinear factorization it decomposes cross-section
into parton density functions characterizing incoming hadrons $\phi(\xi,k_\perp)$   
  at  fixed transverse momentum,  and perturbatively calculable matrix
elements $\widehat\sigma$. 
However, it resumes apart from large logarithms of hard scale also large logarithms coming from 
energy ordering. The formula for high energy factorization while applied to considered here process assumes form:  
\begin{equation}
 \sigma
 = \sum_a  \int  \     
d \xi_1 \ d \xi_2  \  d^2 k_T \ 
 \phi_{a/A} (\xi_1,\mu^2)  \  \widehat\sigma  ( \xi_1 \xi_2 S , Q_T,k_T  , \varphi )  \ 
\phi_{g^*/B}  (\xi_2,k_T,\mu^2)     \;\;  
\label{eq:ktfactform}
\end{equation}	
where as example we took total cross section and where sum runs over quark flavors.
In high-energy factorization framework the parton densities are solutions to
integro-differential evolution equations summing up perturbative terms with
strong ordering condition in rapidity or angle of subsequently emitted partons. Such equations should be
supplemented with some nonperturbative input distribution at initial value of
ordering parameter which then is evolved
with the evolution equation towards larger value of ordering parameter. 
The matrix elements relevant for high energy factorization describe hard
subprocess where at least one of incoming partons is off mass shell. They are
calculated by applying to scattering
amplitudes ${\cal M}$ the high-energy eikonal
projectors.
In reference \cite{Deak:2009xt}  matrix elements relevant for forward
jets phenomenology have been calculated, in fully exclusive form. 
\section{Forward jet phenomenology at the LHC }
We calculate forward jet cross sections  for a typical experimental scenario at LHC.  We require at least two jets with E$_\perp\!>\!10$~GeV, 
where one jet has to be detected in the central region defined by $|\eta_c|\!<2 $ and the other jet is reconstructed in the forward 
region defined by $3\!<\!|\eta|\!<5\!$. 
The jets are reconstructed using the invariant $anti-k_t$-algorithm. 
We compare predictions from running the CASCADE Monte Carlo event generator with the PYTHIA~\cite{Sjostrand:2006za} Monte Carlo event generator running in two 
modes: with and without multi-parton interactions. Both Monte Carlo 
generators 
simulate higher order QCD corrections with parton showers: CASCADE uses parton showers according to the CCFM evolution equation whereas PYTHIA uses DGLAP 
based parton showers.

\subsection{Transverse momentum spectra}
\label{sec:trv}
\begin{figure}[t!]
  \begin{picture}(30,30)
    \put(-40, -210){
      \includegraphics{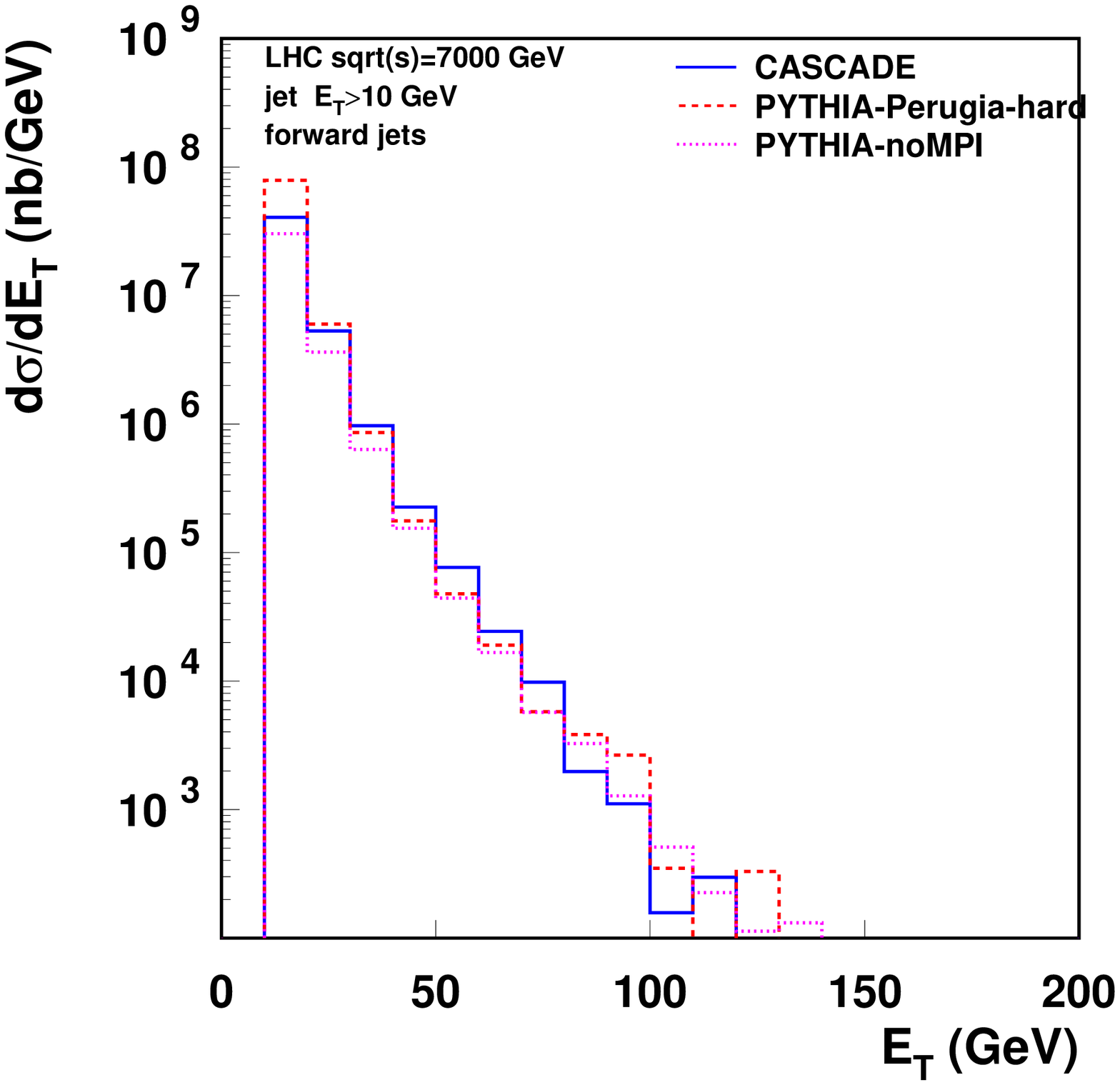}
    }
    \put(220, -210){
      \includegraphics{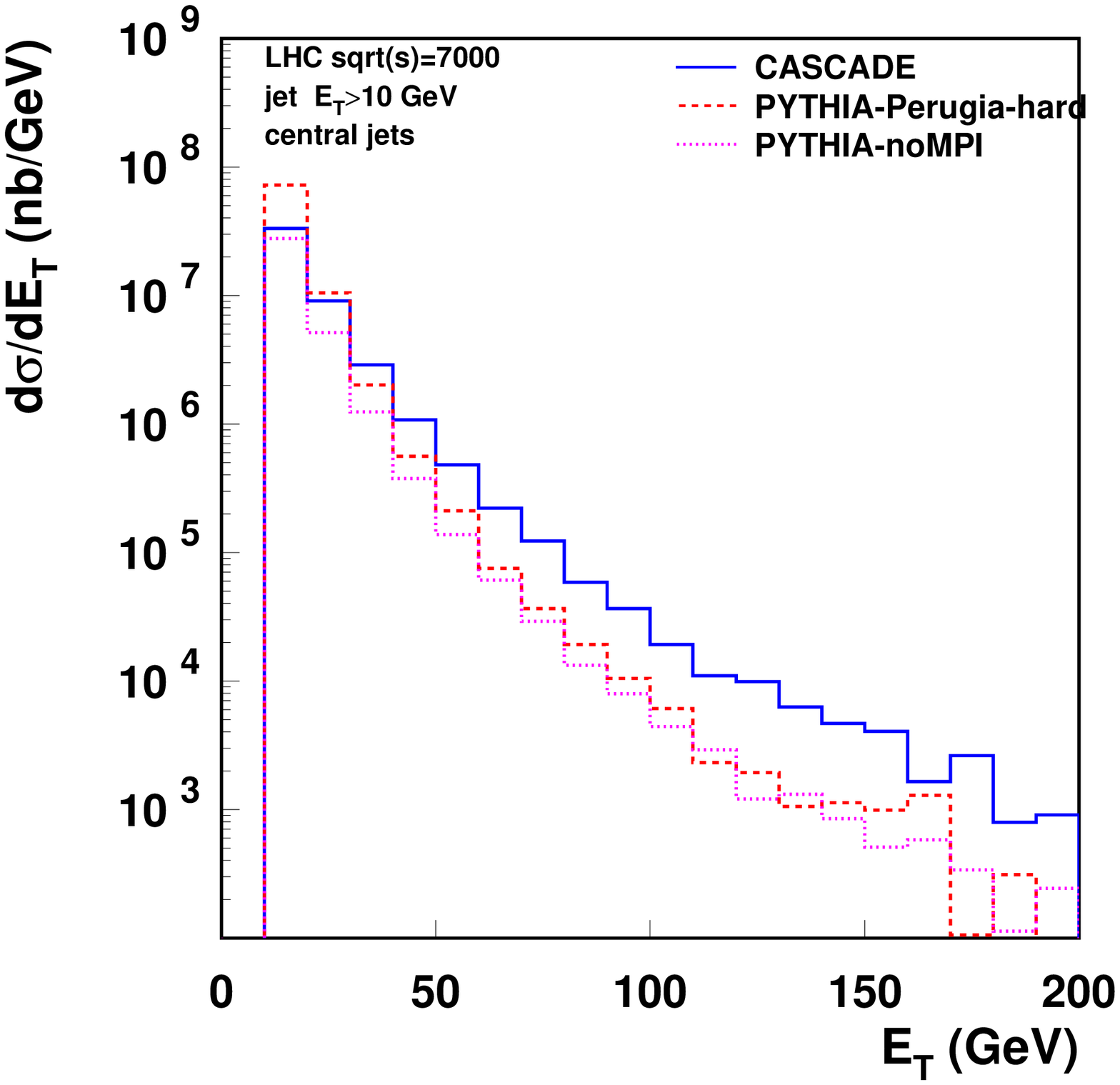}
    }

     \end{picture}
\vspace{6.5cm}
\caption{\em \small Transversal momentum spectra of produced jets at total collision energy $\sqrt s=7\,TeV$ with requirement that p$_\perp\!>\!10\,GeV$. 
We compare predictions obtained from  CASCADE and PYTHIA running in a multiple interactions mode and no multiple interactions mode. Spectrum of forward jets (left);  
spectrum of central jets (right).}
\label{Fig:transversal}
\end{figure}

In Fig.~\ref{Fig:transversal} the prediction of differential cross section $\frac{d\sigma}{d p_{\perp}}$
is shown as obtained from CASCADE and PYTHIA. The cross sections predicted from both simulations at low momentum are of the similar order,  
however, at larger transverse momentum the  CASCADE predicts a 
larger cross section what is clearly visible for central jets (Fig.~\ref{Fig:transversal} right).
This behavior is expected since  CASCADE uses matrix elements which are calculated within high energy factorization scheme allowing for harder transversal
momentum dependence as compared to collinear factorization. Moreover CASCADE 
applies CCFM parton shower utilizing angle dependent evolution kernel which at small $x$ does not lead to ordering in transverse momentum, 
and thus allow for more hard radiations during evolution as compared to based on leading order DGLAP splitting functions Monte Carlo generator PYTHIA. 
The parton shower has major influence on the side where the small $x$ gluon enters the hard interaction, thus the jets in the central region are 
mainly affected by the parton shower. 

\vskip 0.3 cm 
\subsection{Rapidity dependence}
\label{sec:rap}
\begin{figure}[t!]
  \begin{picture}(30,30)
    \put(-40, -190){
      \includegraphics{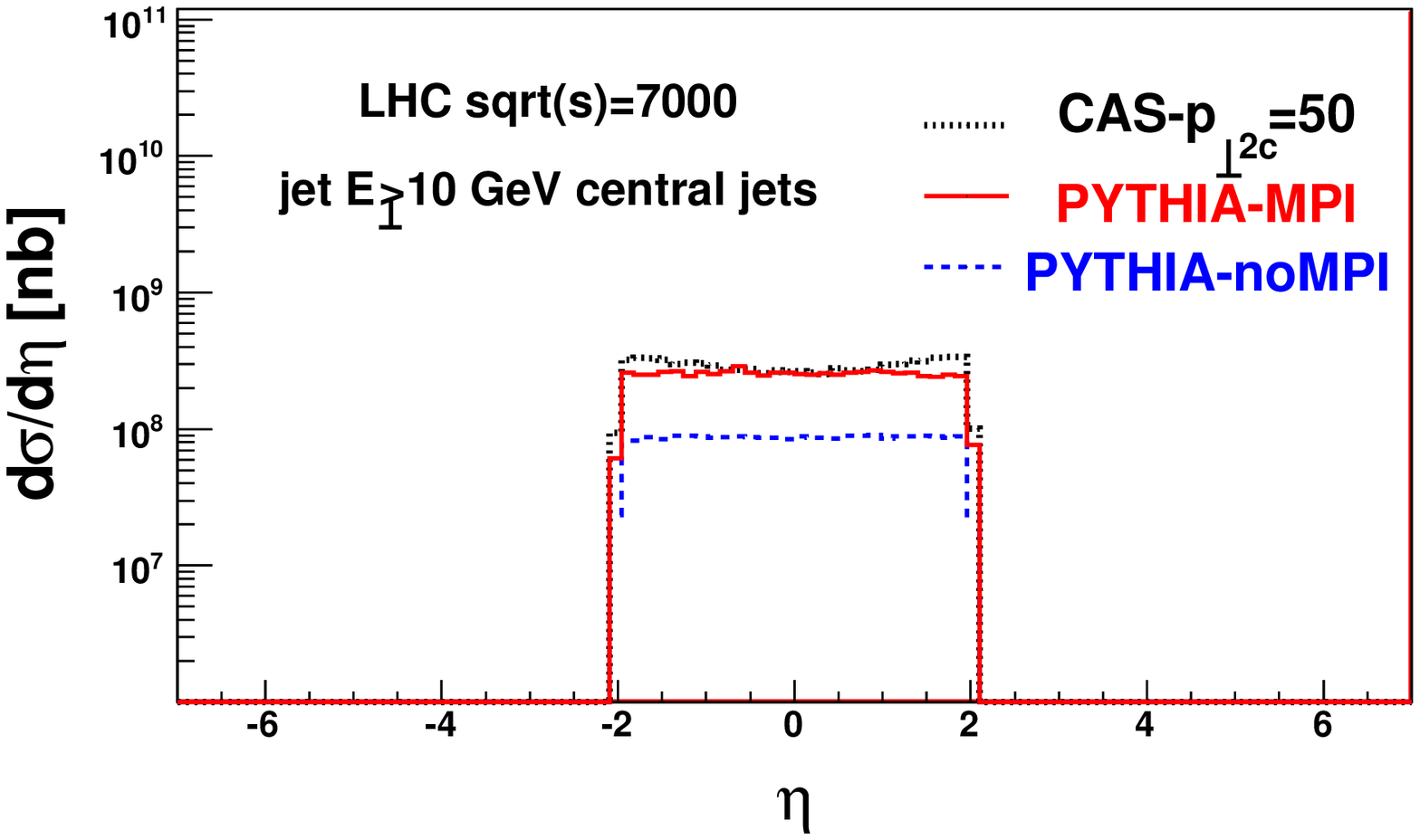}
    }
    \put(220, -190){
      \includegraphics{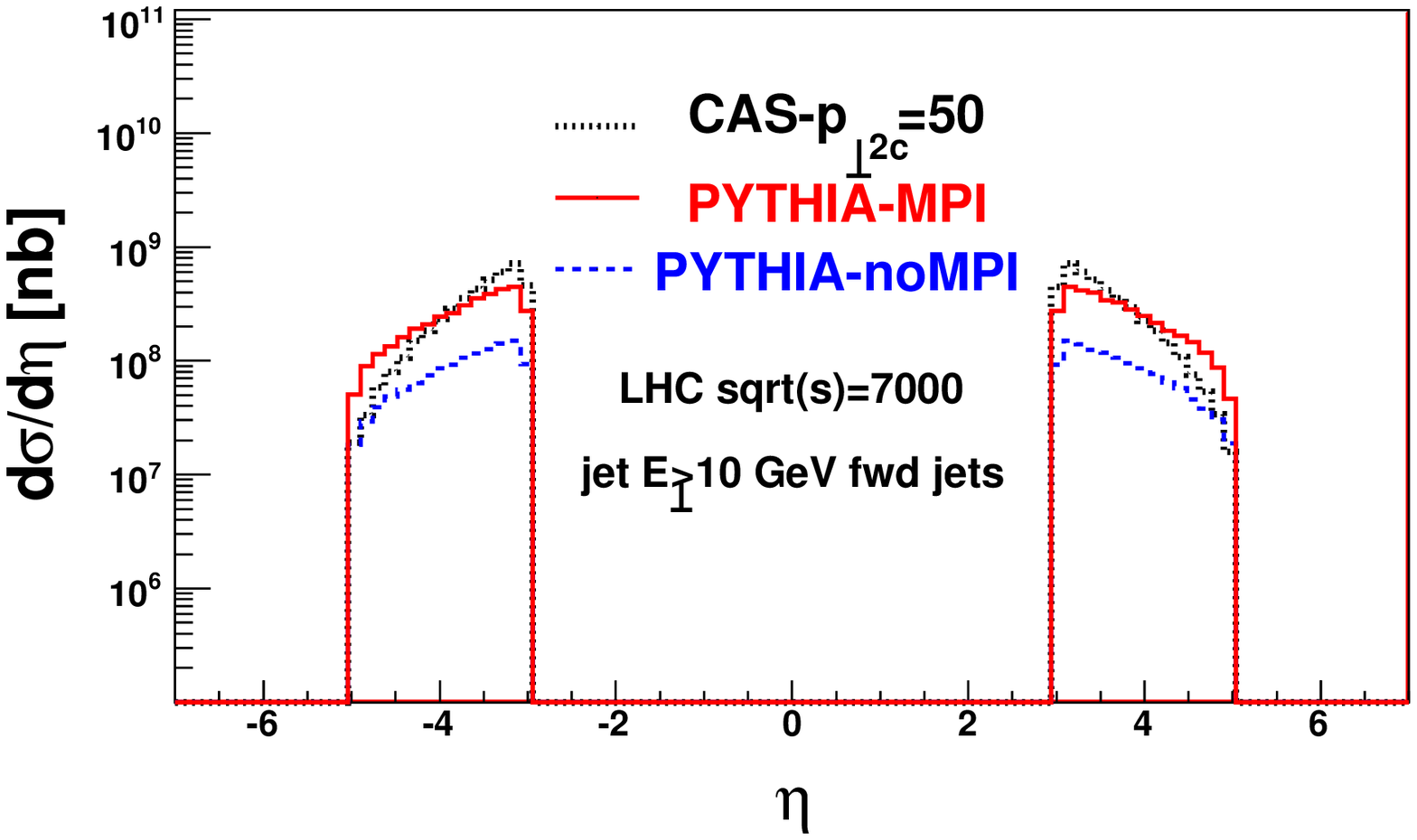}
    }

     \end{picture}
\vspace{6.2cm}
\caption{\em \small  Pseudorapidity spectra of produced jets at total collision energy $\sqrt s=7\,TeV$with requirement that $p_T\!>\!10GeV$. 
We compare predictions obtained from CASCADE and PYTHIA running in multiple interactions mode and no multiple interactions mode. Spectrum of forward jets
(left); spectrum of central jets (right).}
\label{Fig:rapidity}
\end{figure}

In fig.~\ref{Fig:rapidity} we show prediction for pseudorapidity dependence of the cross section in two regions $0\!<\!|\eta|\!<\!2$ and $3\!<\!|\eta|\!<\!5$.
We see that results from CASCADE interpolate between PYTHIA with multiple interactions in the central region and PYTHIA without multiple interactions 
in the forward region.  
The result is due to the fact that  CASCADE (because of angular ordering), and PYTHIA with multiple interactions (because of multi chain exchanges), predict 
more hadronic activity  in the central rapidity region as compared to the collinear shower. 
In the remaining rapidity region cascade uses collinear parton shower of a similar type as in PYTHIA without multiple interactions.

\vskip 0.3 cm

\section*{Acknowledgements}
I would like to thank organizers for a very nice meeting. The results presented in this article have been obtained in collaboration with M. De\'ak, H. Jung, 
F. Hautmann.

\end{document}